\documentclass[aps,superscriptaddress,reprint]{revtex4-2}

\usepackage[utf8]{inputenc}
\usepackage{xcolor}
\usepackage{pst-node}
\usepackage{graphicx}
\usepackage{amsmath, amssymb, amsthm}
\usepackage{mathtools}
\usepackage{hyperref}
\usepackage[shortlabels]{enumitem}
\usepackage{dsfont}
\usepackage{amsthm}
\usepackage{amssymb}
\usepackage{amstext}
\usepackage{amsfonts}
\usepackage{nicefrac}
\usepackage{extarrows} 
\usepackage{graphicx}
\usepackage{relsize}
\usepackage{soul}

\newcommand{\ket}[1]{\vert #1 \rangle}

\newcommand{\calB}{{\mathcal B}}

\newcommand{\R}{\mathbb R}

\begin{document}
\title{Global fermionic mode optimization via swap gates}

\author{Gero Friesecke}%
 \email{gf@ma.tum.de}
\affiliation{%
  Department of Mathematics, Technical University of Munich, Germany
}%

\author{Mikl\'os Antal Werner}
\affiliation{Strongly Correlated Systems Lend\"ulet Research Group, Wigner Research Centre for Physics, H-1525, Budapest, Hungary}

\author{Kornél Kap\'as}
\affiliation{Strongly Correlated Systems Lend\"ulet Research Group, Wigner Research Centre for Physics, H-1525, Budapest, Hungary}
\affiliation{%
Department of Theoretical Physics, Institute of Physics, Budapest University of Technology and Economics, M\H uegyetem rkp. 3, H-1111 Budapest, Hungary
}

\author{Andor Menczer}
\affiliation{Strongly Correlated Systems Lend\"ulet Research Group, Wigner Research Centre for Physics, H-1525, Budapest, Hungary}

\author{\"Ors Legeza}
\email{legeza.ors@wigner.hu }
\affiliation{Strongly Correlated Systems Lend\"ulet Research Group, Wigner Research Centre for Physics, H-1525, Budapest, Hungary}
\affiliation{Institute for Advanced Study, Technical University of Munich,
Germany, Lichtenbergstrasse 2a, 85748 Garching, Germany}
\affiliation{Parmenides Stiftung, Hindenburgstr. 15, 82343, Pöcking Germany}

\date{\today}

\begin{abstract}
An optimal choice of single particle modes leads to an optimal tensor network-based compression of the many-body wave function for systems of indistinguishable particles, which compression can be essential for large, close-to-critical systems. The proposed mode optimization relies on the minimization of the block entropy area, a global quantity that measures entanglement of the wave function along the one-dimensional chain of modes. Extension of the two-site DMRG algorithm by nearest-neighbor mode optimizations is straightforward, but it performs poorly without systematic reordering of modes. Moreover, finding a stationary optimal point in the combination of the fixed-rank MPS manifold and the unitary group of mode rotations requires optimization for every generator, i.e., for every two-mode pair. Here, a systematic joint optimization protocol over the MPS manifold and the full unitary group is presented in which every two-mode pair is accessed by a carefully constructed sequence of two-qubit operations, including swap-gate controlled permutations.
Large-scale DMRG simulations of strongly correlated two-dimensional fermionic lattice models and the multireference Fe$_4$S$_4$ transition metal cluster demonstrate rapid convergence towards the optimum, achieving low energy and low entanglement.  
\end{abstract}

\maketitle

\emph{Introduction:}
Finding an optimal representation of a quantum many-body wave function, i.e., a parametrization with the minimum number of parameters for a given error margin, is a task of utmost importance in modern quantum physics and chemistry~\cite{Bartlett-2007, Carleo-2017, Nagy-2019, Evangelista-2019,Choo-2020}.
Tensor network state (TNS) methods ~\cite{Schollwock-2005,Schollwock-2011,Verstraete-2023,Orus-2019} building on the seminal work of S.~R.~White~\cite{White-1992a} provide an efficient tool to reduce the computational cost to a polynomial form for quantum states with low enough entanglement~\cite{Eisert-2010, Stoudenmire-2012}. The density matrix renormalization group (DMRG) algorithm~\cite{White-1992a, Schollwock-2005, Schollwock-2011} optimizes iteratively the matrix product state (MPS) form of the many-body wave function, where each matrix corresponds to one of the $N$ elementary subsystems that can represent a spin, a qubit, or a single-particle mode. The wave function is compressed by the singular value decomposition-based truncation of matrices, either by defining a fixed maximal matrix rank or keeping the so-called truncation error below a given threshold~\cite{Legeza-2003a}. 

However, the standard DMRG is far from optimal for systems of indistinguishable particles, as it does not exploit the large freedom in choosing the single-particle basis that can lead to an optimally compact wave function representation. Algorithms for mode optimization in TNS algorithms based on local mode-rotations and heuristic reorderings~\cite{Krumnow-2016, Krumnow-2019, Krumnow-2021, Mate-2022, Menczer-2023d}, or based on direct energy minimization~\cite{Murg-2010a} have been developed in the past decade, but they either lack the desired systematicity or are computationally too expensive. 
The usual two-site DMRG algorithm can be straightforwardly extended by local mode-optimization, i.e., nearest neighbor unitary transformations~\cite{Krumnow-2016}; however, a global reordering of the modes based on the rather heuristic Fiedler-vector method~\cite{Barcza-2011} is necessary after several sweeps, in practice, otherwise the algorithm gets easily stuck far away from optimum. However, this reordering protocol does not guarantee that all pairwise mode transformations are accessed, as would be necessary to achieve a stationary point on the full unitary group, and leads to undesirable fluctuations.
On the other hand, when global, energy gradient based optimization protocols over the unitary group are applied~\cite{Murg-2010a}, for example, by replacing the CI solver in the complete active space self consistent field (CASSCF) procedure with DMRG~\cite{Zgid-2008c}, first it is not clear wether the entanglement of the MPS is reduced and, second, in case of strong correlations, a very high accuracy demand is usually necessary for the DMRG to calculate the one- and two-particle reduced density matrices for many intermediate steps~\cite{Legeza-2025}, which makes this procedure very costly.

In this letter, we present a new approach that intertwines minimization over the MPS parameters of the energy with optimization over the unitary group. For the latter, we minimize an entanglement-based global quantity, the \textit{block entropy area} (BEA), defined as the sum of the entanglement entropies over all bonds. The required matrix ranks at a given error margin are closely related to the bond entropies; consequently, BEA quantifies the achievable compactness of the MPS wave function. After the definition of BEA and the discussion of its stationarity condition, we introduce our optimization strategy based on nearest neighbor rotations and swap-gate controlled permutations. 
Theoretical analysis of the method is followed by large-scale DMRG simulations that demonstrate robust convergence during macro-steps, simultaneously achieving low energy and low entanglement with better accuracy if compared to the Fiedler-vector approach. 

\emph{Block entropy area}: 
Fermionic wave functions belonging to the Fock space over $N$ single-particle modes $\varphi_1,...,\varphi_N$ are represented by their coefficient tensor $C(\mu_1,...,\mu_N)$ in the expansion 
\begin{equation}
 \Psi = \sum_{\mu_1,...,\mu_N=0}^1 C(\mu_1,...,\mu_N)\Phi_{\mu_1,...,\mu_N}
\end{equation}
where the many-body basis functions are Slater determinants in the  occupation representation,  
$\Phi_{\mu_1...\mu_N} := |\varphi_{i_1}...\varphi_{i_n}\rangle$  where $\mu_{i}=1$ exactly when $i\in\{i_1,...,i_n\}$, $i_1<\ldots<i_n$. 
The BEA is then defined as
\begin{equation}
   \calB_{\alpha}(C) = \sum_{\ell=1}^{N-1} S_\alpha(\rho_{1,2,...,\ell})
   \label{eq:bea}
\end{equation}
where $\rho_{1,2,...,\ell}$ is the reduced density operator of the first $\ell$ modes, 
\begin{align}
  &\rho_{1,2,...,\ell}(\mu_1,..,\mu_\ell;\mu'_1,...,\mu'_\ell) \nonumber \\
  &=\!\! \sum_{\mu_{\ell+1},...,\mu_N} \!\! C(\mu_1,...,\mu_N) C^*(\mu'_1,...,\mu'_\ell,\mu_{\ell+1},...,\mu_N)
\end{align}
and $S_{\alpha} = \frac{1}{1-\alpha} \ln(\rm{Tr} \rho^\alpha)$ is the R\'enyi entropy of parameter $\alpha$. 
In practice, we set $\alpha=1/2$, but one can also use other concave entropies~\cite{Bengtsson-2017}, like the $S_{\alpha \rightarrow 1}$ von Neumann entropy.
The block entropy area strongly depends on the chosen basis, $\lbrace \varphi_i \rbrace$, whose optimization is the main goal of this work. Due to concavity of $S_{\alpha}$, a low $\mathcal{B}_{\alpha}(C)$ is possible only if the density matrices $\rho_{1,2,...,\ell}$ have many eigenvalues close to zero, i.e., they have good low-rank approximations, where the small eigenvalues are truncated. Consequently, the rank of the $\underline{\underline{A}}^{[i]}(\mu_i)$ matrices in the MPS wavefunction
\begin{align}
    &C(\mu_1,...,\mu_N) = \underline{A}^{[1]}(\mu_1) \underline{\underline{A}}^{[2]}(\mu_2) \underline{\underline{A}}^{[3]}(\mu_3) \dots \underline{A}^{[N]}(\mu_N)
\end{align}
can be kept small.

\emph{Fermionic mode optimization}: Under a single particle unitary mode transformation $U\in U(N)$ which yields the new modes $\varphi_i' = \sum_j U_{ij} \varphi_j$, the coefficient tensor $C$ of a given fixed wavefunction $\Psi$ transforms to $C'=G(U)\, C$ where $G(U)$ is a unitary transformation on the space of many-body coefficient tensors. For the time reversal symmetric case, studied in this work, $C$ and the $\varphi_i$ can be taken to be real-valued, and mode transformations are given by an orthogonal matrix $U\in SO(N)$ after discarding an immaterial overall sign factor. The generalization of the following construction to the full $U(N)$ group is, however, straightforward. 
We aim to minimize the cost function $f(U) = \mathcal{B}_{1/2}(G(U) C)$ over unitary matrices $U$. Such matrices can be parametrized around an arbitrary fixed $U_* \in SO(N)$ as $U=e^A U_*$ with $A$ being real and skew-symmetric. The parametrization is unique for $U$ close to $U_*$. Stationarity of the scalar function $f$ on $SO(N)$ at $U_*$ is equivalent to 
\begin{equation} \label{eq:stat}
    0 = \frac{d}{dt}\Big|_{t=0} f(e^{tA} U_*), ~~ \forall A^T = - A 
\end{equation} 
We notice that any skew matrix $A$ can be expanded over the antisymmetric basis $E_{ij}=e_ie_j^T - e_je_i^T$, where $e_i$ is the unit vector of $\R^N$ whose $i$-th component is $1$ and whose other components are zero. These basis matrices parameterize pairwise rotations like $U_{ij}(\theta) = w^{\theta E_{ij}}$, which corresponds to the mode transformation $\varphi_i'=\cos\theta \varphi_i + \sin\theta \varphi_j$, $\varphi_j'=-\sin\theta \varphi_i + \cos\theta \varphi_j$ while it leaves all other modes invariant. As $E_{ij}$ is the basis of skew matrices, \eqref{eq:stat} is satisfied if and only if $\frac{d}{d\theta} f(U_{ij}(\theta)U_*)|_{\theta=0}=0$ for all $i<j$.
\begin{figure}
    \centering
    \includegraphics[width=0.35
    \textwidth]{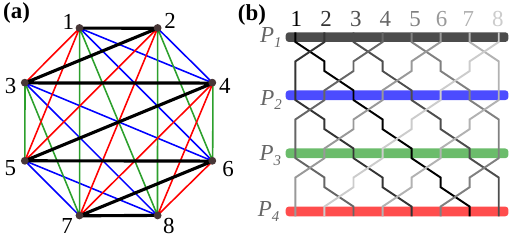}
    \caption{\textit{(a)} Walecki's construction~\cite{Lucas-1882} for $N=8$ modes. The first permutation of modes forms a zig-zag line (with black) over the vertices of the regular polygon with $N$ vertices ($N$ is even). Other permutations (blue, green, and red lines) are generated by rotating the zig-zag line around the center by $2\pi/N$. \textit{(b)} Realization of permutations by swapping nearest neighbor modes in a checkerboard pattern. At every second layer of swap gates, the next permutation of panel \textit{(a)} is formed. For better readability, we used the same colors to mark permutations in the two panels. 
    }    
    \label{fig:permutation}
\end{figure}

\begin{figure*}
    \centerline{     
    \includegraphics[width=1.0\textwidth]{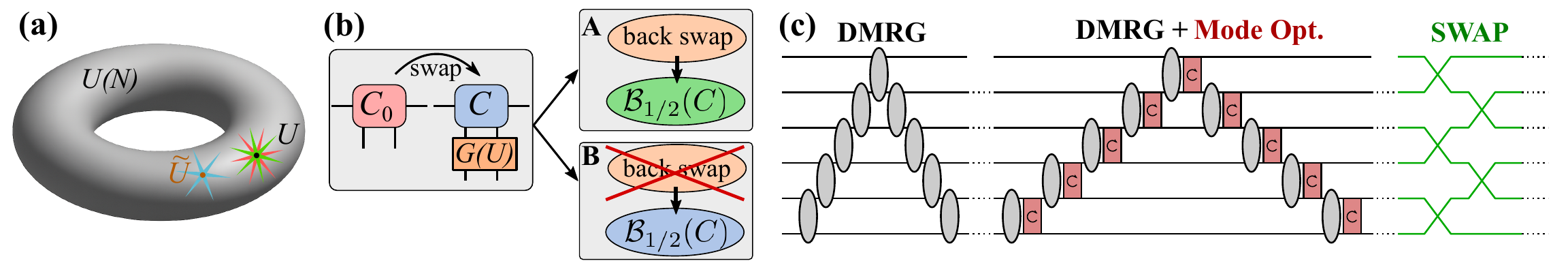}
    }
    \caption{\textit{(a)} Schematic plot of the unitary group $U(N)$. The red star indicates directions reachable from $U$ by optimization of nearest neighbor unitaries, while the green star shows directions accessible by rotations of another permutation. However, a permutation also changes $U$ to $\tilde{U}$, i.e., without employing an inverse permutation, the blue star gets accessible.  \textit{(b)} The global mode optimization algorithm. First, the modes are permuted by swap layers, then nearest neighbor rotations are optimized. In the rigorous "A" algorithm, the inverse permutation would be necessary before cost calculation. Due to its high computational cost, this step is omitted in the practical "B" implementation. \textit{(c)} A macro-iteration step. First, several sweeps of standard DMRG are performed, then sweeps of local mode optimization are executed, and finally, two layers of swap gates generate the next permutation.  
    }
\label{fig:swap}
\end{figure*}
\emph{Local mode optimization and block entropy area}:
First, focus on the nearest neighbor rotations, i.e., on local mode optimization. Here the unitary $U$, under which Hamiltonian transforms as $H(U) = G(U)^\dagger H G(U)$ 
is constructed iteratively from two-mode unitary operators $U_{l,l+1}(\theta)$ while sweeping through the network for $l=1\ldots N-1$~\cite{Krumnow-2016}.
At the micro-iteration step searching $U_{l,l+1}(\theta)$, the half-R\'enyi block entropy $S_{1/2}(\rho_{\{1,2,\dots,l\}})$ is minimized as a function of $\theta$. For iterative optimization protocols, it is usually required for individual optimization steps that they should change only one corresponding local term of the cost function. In our case, as $S_{1/2}(\rho_{\{1,2,\dots,l\}})$ is the only term in \eqref{eq:bea} that is affected by $U_{l,l+1}(\theta)$, i.e., the local optimization fulfills the above criterion, and it necessarily decreases the total cost function $\mathcal{B}_{1/2}(C)$. 
This local optimization procedure, however, converges to a sub-optimal unitary in general (see Fig.~\ref{fig:spinless_6x6_BEA_swap}). This is the consequence of the missing long-range rotations, whose derivative must also vanish. In former works, sub-optimal unitaries have been avoided by heuristic reorderings based on quantum correlations using the Fiedler-vector method~\cite{Barcza-2011,Krumnow-2016}. This latter step provides a new set of pairwise $\theta_{ij}$ components, but it does not traverse through the full $\theta_{ij}$ parameter space, and leads to undamped fluctuations once the optimum is approached. In the present work, a systematic reordering scheme is presented (see Fig.~\ref{fig:permutation})  that guarantees the coverage of all mode-pairs, is numerically cheaper, and, most importantly, provides better optimal modes.

\emph{Swap-gate controlled permutations and nearest neighbor rotations}: 
Direct optimization of long-range pairwise rotations based on \eqref{eq:stat} alongside energy minimization would need a significant modification of DMRG. However, by employing $N/2$ appropriate re-orderings of the modes,  all pairwise rotations $U_{ij}(\theta)$ can be realized in nearest neighbor positions. 
Such an optimal set of permutations $\tau_1,...,\tau_{N/2}$ can be generated by Walecki's method~\cite{Lucas-1882} for even $N$'s, which was originally developed to get the Hamiltonian decomposition of complete graphs.~\cite{Bermond-1978, Wilson-1996}. In the method, summarized in Fig.~\ref{fig:permutation}, a regular polygon of $N$ vertices is drawn. Permutations correspond to zig-zag lines in the polygon, which are transformed to each other by rotations around the center. It is easy to see that the $\frac{N}{2}$ zig-zag lines cover all the edges of the complete graph of $N$ vertices (modes) exactly once, thus every pair of modes is in nearest neighbor position once during the process. In case of an odd $N$, we simply generate the permutations for $N+1$ and then erase the extra mode from the permutations. In that case, some mode pairs are accessed more than once during the process.

A pleasant feature of Walecki's method from the algorithmic point of view is the fact that consecutive permutations are easily generated by two layers of nearest-neighbor swap operations placed in a checkerboard pattern. (see Fig.~\ref{fig:permutation}b). In the language of matrix product states, these swap operations can easily be executed by the elementary step of the time-evolving block decimation (TEBD) algorithm~\cite{Vidal-2003a}, hence transforming the state vector between consecutive permutations is simple, numerically cheap, and thus one can spare several initialization sweeps in the next macro iteration.
In order to avoid truncation of the wavefunction, the bond dimension has to be increased by a factor of $q$ for both the forward and the backward sweeps, where, in our case,  $q = 2$ is the local dimension of a spinless fermionic mode.

An undesired effect of the above-introduced reordering scheme is that it also changes the unitary $U$ itself, which can result in an increased value of the cost $\mathcal{B}_{1/2}(C)$ (see Fig.~\ref{fig:swap}). Following \eqref{eq:stat} rigorously, the cost should be calculated for the original order, i.e., modes should be back-ordered in all steps when $\mathcal{B}_{1/2}(C)$ is calculated (see Fig.~\ref{fig:swap}). However, this would increase the computational cost to an unacceptable extent. To circumvent this problem, we use the less precise but more practical protocol, where $\mathcal{B}_{1/2}(C)$ is always measured for the actual permutation. As our results show, robust convergence can be achieved to a mode set that is superior to that of the heuristic Fiedler-vector method. As a possible explanation, we add that reordering by two layers of swap gates is local in the sense that the maximal distance between sites that have been nearest neighbors is five. The following local mode optimization steps have, therefore, a good chance to remove an undesired increase in $\mathcal{B}_\alpha(C)$, but this is not mathematically guaranteed.

\textit{Global DMRG-mode-optimization protocol:}
During optimization, consecutive macrosteps are executed until convergence. Their required number roughly equals $N$.
In one macrostep, one first performs $N_{\rm SW}^{\rm dmrg}$ number of standard DMRG sweeps, where only the energy is optimized, then $N_{\rm SW}^{\rm opt}$ local mode optimization sweeps, where both energy and $\mathcal{B}_{1/2}(C)$ are optimized, and finally two layers of swap gates are applied to prepare the next permutation. (For further details, see the Supplemental Material~\cite{suppmat}.) 
In the last step, one can temporarily increase the bond dimension by $q^2$ to avoid harming the wave function, and truncate it back in the next macrostep; however, this has only a minor effect.

\vskip 0.1cm
\emph{Numerical results:}
Numerical simulations will be presented for a very general form of the Hamiltonian operator, 
\begin{equation}
\mathcal{H} = \sum_{ij\alpha\beta} T_{ij}^{\alpha\beta} 
                c^\dagger_{i\alpha}c_{j\beta} +
                \sum_{ijkl\alpha\beta\gamma\delta} 
      {V_{ijkl}^{\alpha\beta\gamma\delta}
      c^\dagger_{i\alpha}c^\dagger_{j\beta}c_{k\gamma}c_{l\delta}},
\label{eq:ham}
\end{equation}
implemented in our code~\cite{dmrg-budapest},  
that can treat any form of non-local interactions related to two-particle scattering processes. In Eq.~\ref{eq:ham}
$i,j,k$ and $l$ label the modes in the tensor network, while $\alpha,\beta,\gamma$ and $\delta$ denote spin indices. It can thus describe nuclear shell problems, or quantum chemistry, even in the relativistic domain~\cite{White-1992b,Xiang-1996,White-1999,Schollwock-2005,Noack-2005,Chan-2008,Schollwock-2011,Knecht-2014,Dukelsky-2004,
Orus-2014,Legeza-2015,Szalay-2015a,Legeza-2018a,Shapir-2019,Barcza-2021b,Baiardi-2020}.
By proper choices of $T_{ij}^{\alpha \beta}$ and $V_{ijkl}^{\alpha\beta\gamma\delta}$, the Hamiltonian can also describe lattice models of fermions with local interactions in two spatial dimensions with periodic boundary conditions~\cite{Krumnow-2021, Menczer-2023d}.
However, the local nature of the Hamiltonian is lost after mode optimization, and the generic form \eqref{eq:ham} is recovered. In this work, we demonstrate the method for the model of spinless fermions on a two-dimensional square lattice with hoppings of strength $t$ and $t'$ to nearest and next-nearest neighbors, respectively,  and nearest neighbor interactions $V$~\cite{Menczer-2023d}. The chosen parameters $t'/t = 0.4$ and $V/t = 0.4$ are near a critical point~\cite{Menczer-2023d}. Results for the two-dimensional spinful Hubbard model are also shown in the Supplemental Material for both attractive and repulsive interactions.

\begin{figure}
    \centering     
    \includegraphics[width=0.48\textwidth]{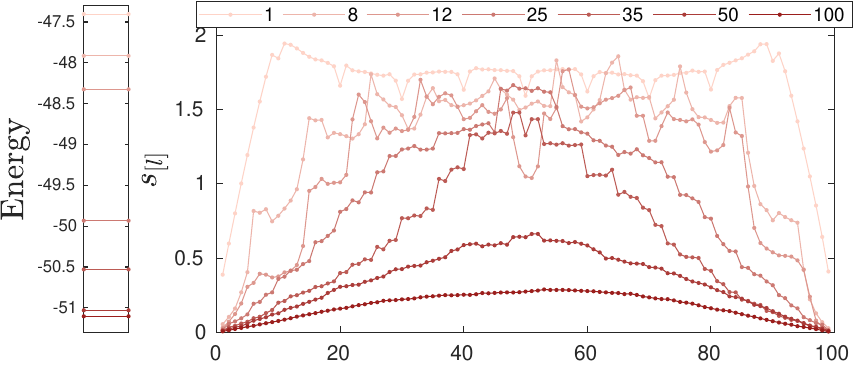}    
    \caption{
    Ground state energy (left) and block entropy, $S_1(\rho_{1,2,\dots,\ell})$, as a function of the left block size, $\ell$ (right), for some selected mode transformation macro iteration cycles indicated at the top of the figure,
    with fixed bond dimension $D_{\rm mo}=80$ and 11 sweeps per macro iteration $N^{\rm{dmrg}}_{\rm{SW}}=2$, $N^{\rm{opt}}_{\rm{SW}}=7$ for the half-filled $10\times 10$ two-dimensional spinless fermionic model with $t'/t=0.4$ and $V/t=0.8$. The area under each curve corresponds to the block entropy area $\calB_1(C)$ shown in Fig.~\ref{fig:spinless_6x6_BEA_swap}.
    Bringing each mode pair to neighbor positions at least once needs $N/2=50$ permutations, i.e., 50 mode transformation macro iterations.
    }  
    \label{fig:spinless_10x10_S_swap}
\end{figure}
The initial ordering of the modes in the 2D lattice follows the usual snake-path~\cite{Liang-1994}, from which the series of above-described macro iterations are started. In Fig.~\ref{fig:spinless_10x10_S_swap}, the block entropy profiles are shown in a $10 \times 10$ lattice for various macro iterations at bond dimension $D=80$, and also the corresponding ground state energies are indicated. We observe a systematic convergence to a state with a smooth, symmetric entropy profile and low energy. The necessary number of macro iterations depends on the number of modes. As one needs $N/2$ permutation steps to cover all mode pairs, this also roughly sets the minimal number of macro iterations, but to reach convergence, one rather needs $N$ macro iterations in practice (see also Fig.~\ref{fig:spinless_6x6_BEA_swap}).
The various parts of macro-iterations are better visible for the $6 \times 6$ system analyzed in Fig.~\ref{fig:spinless_6x6_BEA_swap}, where a smaller number of macro-iterations is sufficient. Without mode optimization, the area $\mathcal{B}_1(C)$ is high, and the ground state energy is above the optimal value by $0.6t$. Omitting the swap layers and employing local mode optimization results only in marginally better results. In contrast, we observe a systematic and large decrease both in energy and $\mathcal{B}_1(C)$ once the swap layers are turned on. In the figure, various colors represent the distinct steps of macroiterations: black dashed lines indicate initial DMRG-only sweeps, red dots indicate sweeps with local mode optimizations, while green dots show the two sweeps of swap layers. The DMRG-only sweeps following the swap layers are very important, as they optimize the ground state MPS for the new permutation. We finally note the smooth convergence for the less rigorous but cheap algorithm of Fig.~\ref{fig:swap}b, which is due to the locality of the swap gates and the smooth stationary entropy profile. In the Supplemental Material, comparison with the Fiedler-vector method and additional results on the half-filled Hubbard model with strong repulsion are shown~\cite{suppmat}. There, the degenerate antiferromagnetic Mott insulator ground states result in strong oscillations. Nevertheless, the best basis of the current method results in significantly lower energy also for the Hubbard model, where the Fiedler-vector method shows strong random fluctuations~\cite{suppmat}. Similar results are also shown for the $\mathrm{Fe_4 S_4}$ cluster, which is a numerically hard chemical system due to its strong multi-reference character~\cite{Sharma-2014, suppmat}.

\begin{figure}
    \centering
    \includegraphics[width=0.48\textwidth]{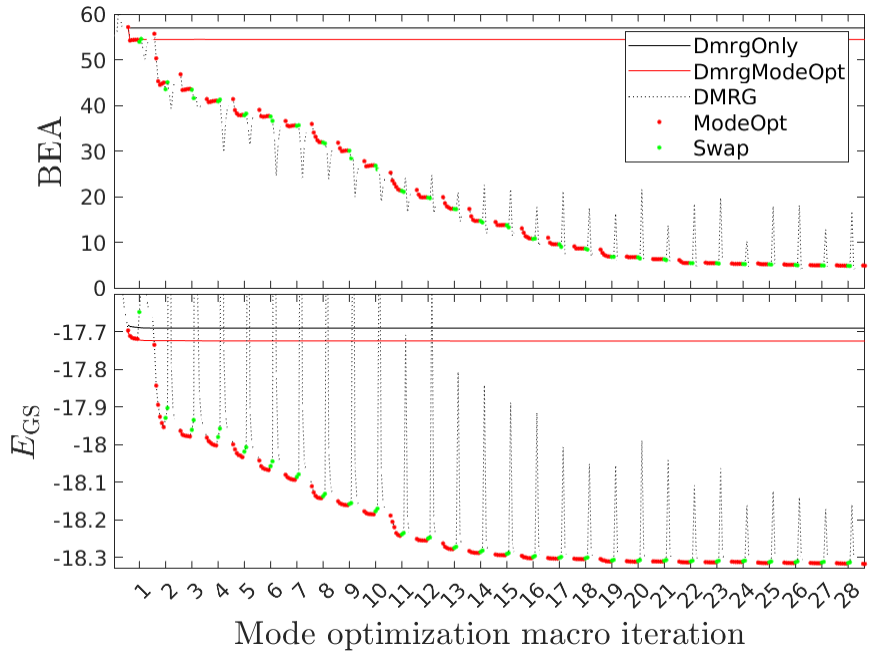}
    \caption{(a) Block entropy area $\calB_1(C)$ and (b) ground state energy for the first 28 mode optimization macro iterations with $D_{\rm opt}=80$, $N_{\rm SW}^{\rm dmrg}=6$, $N_{\rm SW}^{\rm opt}=6$, and 
    for the $6\times 6$ spinless model.
    The dashed line indicates data points obtained at DMRG-only steps, red dots refer to local mode optimization, and green dots to the application of swap gates at the end of each macro iteration.
    For further reference, solid black and red lines show results obtained by DMRG only and by DMRG together with local mode optimization, but without swap gate layers, respectively.
    }    
    \label{fig:spinless_6x6_BEA_swap}
\end{figure}

\vskip 0.1cm
\emph{Conclusion}:
In this work, we proposed a general approach to find an optimal representation of a quantum many-body wave function via global fermionic mode optimization.
The stationary point on a fixed rank matrix product state manifold is obtained via a joint optimization of modes on the Grassmann manifold, together with energy minimization. Swap gates controlled permutations are leveraged to access all generators of the unitary group of single-particle basis rotations. The minimization of the global quantity, the block entropy area, guarantees that the method fulfills all criteria concerning partial derivatives. Numerical results via large-scale density matrix renormalization group simulations on strongly correlated two-dimensional fermionic lattice models, both spinless and spinful, are discussed. The precision of the method is also compared to the formerly used Fiedler-vector-based scheme and is shown to provide superior results~\cite{suppmat}. Boosting the precision of the basis at the low-to-medium bond dimensions used in mode optimization is very important, as an offset in the ground state energy at these low dimensions usually persists partly if one performs a high-$D$ DMRG calculation later, using the optimized basis~\cite{Menczer-2023d}.  

The new protocol is very robust, the convergence is guaranteed by construction, and it is also numerically cheaper as the extraction of two-mode correlations is not necessary for every macro iteration, while the state vector is also preserved between macro iterations. We note that the overall computational complexity and wall time for a given error margin can be reduced by several orders of magnitude when a mode-optimized basis is utilized ~\cite{Menczer-2023d}. 

\vskip 0.1cm
\emph{Acknowledgments:}
This work has been supported by the Hungarian National Research, Development and Innovation Office (NKFIH) through Grant Nos.~K134983 and TKP2021-NVA-04,
by the Quantum Information National Laboratory
of Hungary. \"O.L. acknowledges financial support
by the Hans Fischer Senior Fellowship programme funded by the Technical University of Munich – Institute for Advanced Study and by
the Center for Scalable and Predictive methods for Excitation and Correlated phenomena (SPEC), funded as part of the Computational Chemical Sciences Program  FWP 70942 by the U.S. Department of Energy (DOE), Office of Science, Office of Basic Energy Sciences, Division of Chemical Sciences, Geosciences, and Biosciences at Pacific Northwest National Laboratory.
M.A.W. has also been supported by the Janos
Bolyai Research Scholarship of the Hungarian Academy of Sciences.
The simulations were performed on the national supercomputer HPE Apollo Hawk at the High Performance Computing Center Stuttgart (HLRS) under the grant number MPTNS/44246.

%

\clearpage
\onecolumngrid

\setcounter{page}{1}\renewcommand{\thepage}{S\arabic{page}}
\setcounter{equation}{0}\renewcommand{\theequation}{S\arabic{equation}}
\setcounter{figure}{0}\renewcommand{\thefigure}{S\arabic{figure}}
\setcounter{table}{0}\renewcommand{\thetable}{S\arabic{table}}
\setcounter{section}{0}\renewcommand{\thesection}{S\arabic{section}}

\renewcommand{\citenumfont}[1]{S#1}
\renewcommand{\bibnumfmt}[1]{[S#1]}

\begin{center}
\large{\textbf{Supplemental material to "Global fermionic mode optimization via swap gates"}}
\end{center}
\normalsize

\begin{figure}[!h]
    \centering
    \includegraphics[width=1.0
    \textwidth]{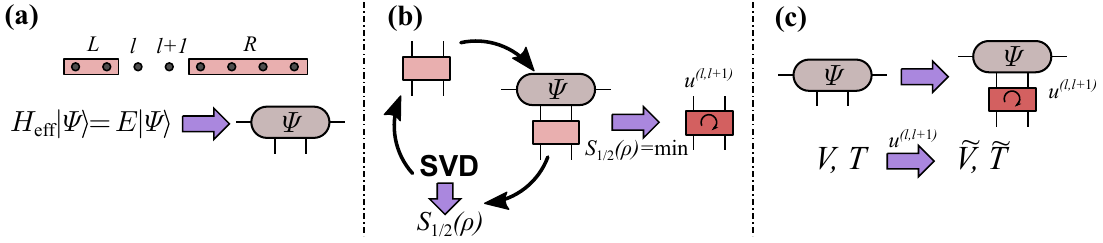}
    \caption{Mode optimization micro iteration at bond position $(l,l+1)$. \textit{(a)} The ground state of the $\hat{H}$ in is numerically determined in the effective Hilbert space $\mathcal{H}_\mathrm{eff}$, similarly to usual DMRG. \textit{(b)} Iterative search for the two-site unitary $u^{(l,l+1)}$ by minimizing the Rényi block entropy $S_{1/2}(\rho)$. An SVD is required in every iteration cycle to calculate $S_{1/2}(\rho)$. \textit{(c)} Tensor form of $\Psi$ is rotated by $u^{(l,l+1)}$ and the Hamiltonian parameters $T$ and $V$ are updated accordingly.
    }    
    \label{suppfig:microiter}
\end{figure}
\section{Algorithmic details}

\subsection{Mode Optimization Micro-Iteration}
We formulate the mode optimization micro-iteration as a 
modification of 
the local optimization procedure of
the Density Matrix Renormalization Group (DMRG) algorithm.
The 
DMRG method provides an iterative algorithm for finding the ground state and the low-lying excited states of the Hamiltonian defined in Eq.(6) of the main text. The many-body wavefunction $\ket{\Psi} = \sum_{\lbrace \mu_1 \dots \mu_N \rbrace} C(\mu_1, \mu_2, \dots \mu_N) \ket{\Phi_{\mu_1 \mu_2 \dots \mu_N}}$ is represented by a matrix product state,
\begin{equation}
    C(\mu_1, \mu_2, \dots \mu_N) = A^{\mu_1}_{[1]}\cdots A^{\mu_N}_{[N]} \; ,
\end{equation}
where the components $A_{[l]}^{\mu_l}$ are (complex-valued) matrices of size $\mathbb{C}^{D_{l-1}\times D_l}$. Here $D_l$ stands for the bond dimension with $D_0 = D_N = 1$. The DMRG algorithm optimizes the components $A_{[l]}$ iteratively by sweeping forward and backward along the MPS chain several times. The two-site variant of DMRG optimizes two neighboring matrices (e.g. $A_{[l]}^{\mu_l}$ and  $A_{[l+1]}^{\mu_{l+1}}$) at the same time in one iteration step. Such a step consists of three main parts. First, the desired eigenstate of the so-called effective Hamiltonian is determined by an iterative eigensolver, then the eigenstate is factorized by the Schmidt decomposition that provides the updated matrices ($\tilde{A}_{[l]}^{\mu_l}$ and  $\tilde{A}_{[l+1]}^{\mu_{l+1}}$). Finally, the sweep position $l$ is moved forward or backward by one site, and the new effective Hamiltonian is formed by the so-called renormalization step, which involves one of the two updated matrices $\tilde{A}$ depending on the direction of the sweep. For a thorough introduction to DMRG, we guide the readers to Refs.~\cite{S.Schollwock-2005,S.Schollwock-2011}.

The main source of error in DMRG is truncation, i.e., the restriction on the bond dimension $D_l$ of the components $A^{{\mu_l}}_{[l]}$. One can introduce a hard cutoff $D_{\max}$ by requiring $D_l \le D_{\max}$ for every component but one can also use adaptive bond dimension by setting a threshold for the discarded entropy~\cite{S.Legeza-2003a}. However, the fact that the amount of entanglement through the bond determines the sufficient bond dimension is independent of the chosen truncation scheme. Reducing the entanglement by the re-definition of single particle modes can lead to more efficient parametrization of the wave function. 

Within the mode optimization micro-iteration (see Supp. Fig.~\ref{suppfig:microiter}), we employ a 
modified DMRG protocol, namely, after the diagonalization step is done, we search the two-site single particle unitary $u^{(l,l+1)}_\sigma$ that mixes the modes of site $l$ and $l+1$ and minimizes the Rényi entropy $S_\alpha(\rho_{1,2,\dots,l})$ at bond $(l,l+1)$. This optimization requires the iterative re-calculation of SVD of the wave function tensor whose run-time can be comparable to the diagonalization step. We also note that, in general, the 2-site unitary can depend on the spin $\sigma$ but one can also enforce spin-independent rotations. After this optimization step, the effective Hamiltonian is updated by $u^{(l,l+1)}$ through the according rotation of the integrals $T_{ij}^{\alpha \beta}$ and $V_{ijkl}^{\alpha \beta \gamma \delta}$, and the micro-iteration is completed by the SVD and renormalization steps of the standard DMRG iteration, as defined above.

Using such nearest-neighbor optimization is motivated by the fact that any global single-particle unitary $U$ can be built by the consecutive action of nearest-neighbor rotations. However, as it is evident from Fig.~4 of the main text, employing such local optimizations only performs poorly compared to the global algorithm, whose details are described below. The reason for this poor performance is simple: the fact that the global unitary can be written as a product of nearest-neighbor unitaries does not mean that it can also be optimized in that form by a 'greedy' algorithm that attempts to reduce the entanglement in every single micro-iteration. Nevertheless, the above-described mode optimization micro-iteration remains an essential step of the global algorithm.

\subsection{Mode Optimization Macro-Iteration}
\begin{figure*}
    \centering
    \includegraphics[width=0.8\textwidth]{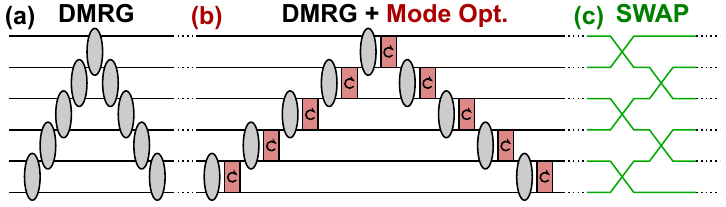}
    \caption{Mode optimization macro iteration. \textit{(a)} Several sweeps of standard DMRG iterations (no mode optimization) are executed to provide an acceptable initial MPS state. \textit{(b)} Sweeps with local state and mode optimization micro iterations (see Fig.~\ref{suppfig:microiter}). The grey ellipses indicate the diagonalization steps that are also part of each micro iteration. \textit{(c)} Before the next macro-iteration step, two layers of swap gates act to form the next permutation. Note that the same coloring is used (grey-red-green) in the numerical figures about $E_{GS}$ and BEA to indicate the different types of steps.
    }    
    \label{suppfig:macroiter}
\end{figure*}

The main goal of the global optimization algorithm is to avoid getting stuck unintentionally at high entanglement due to the locality and greediness of the protocol using nearest neighbor micro-iterations only. Therefore, in the global algorithm, we reorder the modes regularly after several sweeps of micro-iterations. The mode optimization macro-iteration consists, therefore, of three main parts (see Supp. Fig.~\ref{suppfig:macroiter}). First, several sweeps of standard DMRG iterations are performed that produce a sufficiently converged MPS state. Then, sweeps of mode optimization micro-iterations are executed that reduce the entanglement, but usually get stuck after several sweeps. Finally, we reorder the modes by applying swap gates according to Walecki's protocol. As it is also stated in the main text, one nice property of this swap-gate-based reordering is the preservation of the MPS form of the wave function. More precisely, one needs only simplify the micro-iteration step by removing the diagonalization step and by setting $u^{(l,l+1)}$ to the mode-swap or identity operator according to the brick wall pattern. As demonstrated in Figs.~3-4 of the main text, these 
modifications tremendously improve the algorithm's performance. We have to note, however, that the swap-gate layer has unpleasant consequences, once the optimal mode set is approached, because a global algorithm should not only find the optimal modes but also their optimal order. The regular reordering protocol, however, changes the order also in a case when it has already been optimal, and the local micro-iterations cannot entirely remove the unintentionally elevated entropy. Therefore, oscillations of period $N/2$ are observed both in $\mathcal{B}_\alpha(C)$ and $E_{\mathrm{GS}}$. The amplitude of these oscillations depends on the structure of the ground state $\ket{\Psi}$ and, in practice, one can take the mode set from an iteration where $\mathcal{B}_\alpha(C)$ takes its minimal value.

\section{Further numerical results}
\subsection{Comparison with the heuristic Fiedler-vector based method for the 2D spinless model}
In this section, we provide additional data about the spinless model that has also been analyzed in the main text. The precise Hamiltonian of the model is
\begin{equation}
\mathcal{H} = \sum_{\langle i,j\rangle} \left( -t\,c^\dagger_{i}c_{j} + h.c \right) + \sum_{\langle i,j\rangle'} \left( -t'\,c^\dagger_{i}c_{j} + h.c \right) +
               \sum_{\langle i,j \rangle} V n_{i} n_{j}\,,
\label{eq:ham-spinless}
\end{equation}
where $c^\dag_i$ and $c_i$ create and annihilate a fermion on lattice site $i$, $n_i = c^\dag_i c_i$ measures the occupation of the site, and the unprimed/primed sum goes over nearest/next-nearest neighbors.

\begin{figure*}
    \centering
    \includegraphics[width=0.76\textwidth]{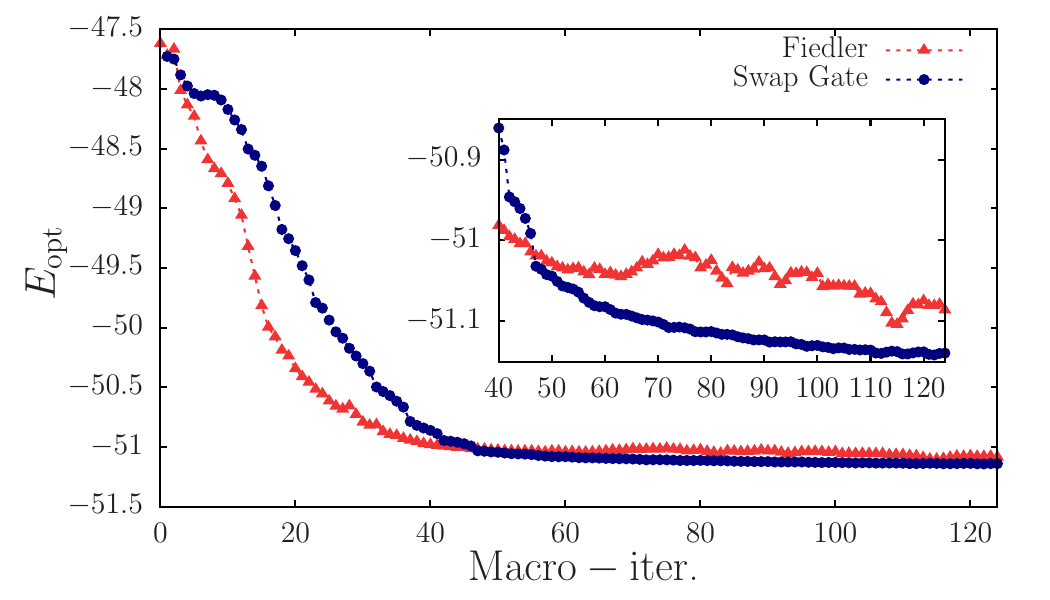}
    \caption{The ground state energy as a function of macro iteration number for the $10 \times 10$ spinless Fermion model \eqref{eq:ham-spinless} for $t'=0.4t$ and $V=0.8t$. Blue circles indicate the swap-gate-based, while red triangles are for the heuristic method. The inset shows the result near the optimum on a finer scale. 
    }    
    \label{suppfig:spinless}
\end{figure*}

In this supplemental analysis, we compare the new, swap-gate-based method to the formerly used protocol using Fiedler vector-based reorderings. In Fig.~\ref{suppfig:spinless}, the convergence of the ground state energy is shown for the macro iteration cycles. In both cases, the bond dimension has been set to $D_{\mathrm{mo}}=80$, and a high number of sweeps per macro iteration, $N_{\mathrm{sweep}}=25$, has been used to ensure temporary convergence of the wave function at each cycle. In summary, both methods show robust convergence in the early macro iterations, with a faster convergence for the heuristic Fiedler method. In contrast, the heuristic method shows strong uncontrolled fluctuations close to the optimum. Moreover, the best energy (optimized basis) of this method within the fluctuating plateau is significantly higher than that of the new method. 
In contrast, the convergence of the ground state energy with the new, swap-gate-based method is smooth and goes below the best Fiedler-vector result in energy by $\Delta E = 0.04t$.

\subsection{2D Hubbard model on square lattice.}
We have also performed simulations analogous to the main text on the 2D Hubbard model,
\begin{equation}
\mathcal{H} = \sum_{\langle i,j\rangle\alpha} \left( -t\,c^\dagger_{i\alpha}c_{j\alpha} + h.c \right) + 
               \sum_i U n_{i\alpha} n_{j\beta}\,,
\label{eq:ham-hubbard}
\end{equation}
where $c^\dag_{i\alpha}$ and $c_{i\alpha}$ are the usual creation and annihilation operators of electrons with spin $\alpha \in \lbrace \uparrow, \downarrow \rbrace$ on lattice site $i$, $\langle i,j \rangle$ indicate nearest neighbor sites,  $n_{i\alpha} = c^\dag_{i\alpha} c_{i\alpha}$, while $t$ and $U$ measure the strength of hopping and interaction. In the following numerical examples, $8 \times 8$ lattices have been considered. For the interaction strength, we have used the same parameters ($t=1$ and $U=\pm4$) as in Ref.~\cite{S.Menczer-2023d}.

\begin{figure*}
    \centering
    \includegraphics[width=0.78\textwidth]{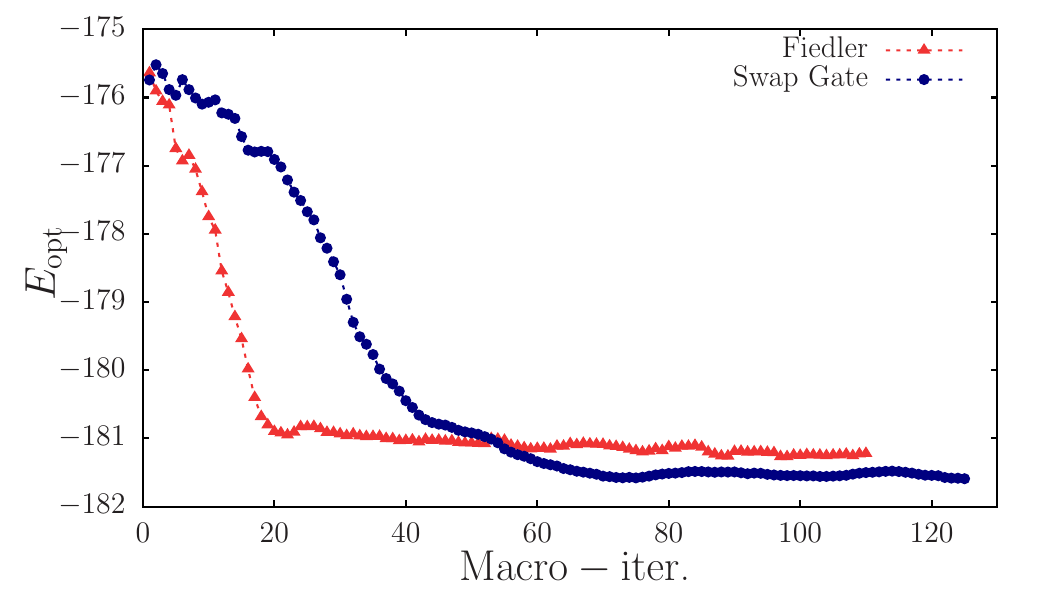}
    \caption{The ground state energy as a function of macro iteration number for the $8 \times 8$ Hubbard model \eqref{eq:ham-hubbard} at $U=-4t$ with $D_\mathrm{opt}=128$. The best energy of the Fiedler-method is $E^\mathrm{Fiedler}_{\mathrm{best}} = -181.27t$, while the swap-gate-based method reaches $E^\mathrm{swap}_{\mathrm{best}} = -181.59t$.
    }    
    \label{suppfig:hubbard_negU}
\end{figure*}
Our central aim in these simulations is to demonstrate the superiority of the new, swap-gate-based protocol over the formerly used, Fiedler-vector-based heuristic reordering~\cite{S.Krumnow-2016,S.Krumnow-2021,S.Menczer-2023d}. First, we repeat the analysis similar to the 2D spinless case for $U=-4t$. In the attractive case, electrons form pairs, and the optimal modes are expected to be the same for $\uparrow$ and $\downarrow$ spins. Consequently, we have enforced the same rotation angles for the two spin components. To make a fair comparison, the sweep number per macro-iteration ($N_{\mathrm{sweep}}=25$), the bond-dimension ($D_{\mathrm{opt}}=128$), and the accuracy threshold of the L\'anczos method were the same in the two cases.
Similarly to the spinless case, we see robust convergence both for the heuristic and the swap-gate-based methods (see Supp. Fig.~\ref{suppfig:hubbard_negU}) in the early macro iterations, with faster convergence for the heuristic Fiedler-vector-based method. Once the optimum is approached, the new swap-gate-based method reaches a significantly lower ground state energy, $\Delta E \approx 0.3t$. We note also that the heuristic method shows uncontrolled fluctuations again, while the swap-gate-based method shows slight oscillations of period $N/2$, i.e., the period where the swap gates reconstruct the mirrored permutation of the original order.

\begin{figure*}
    \centering
    \includegraphics[width=0.78\textwidth]{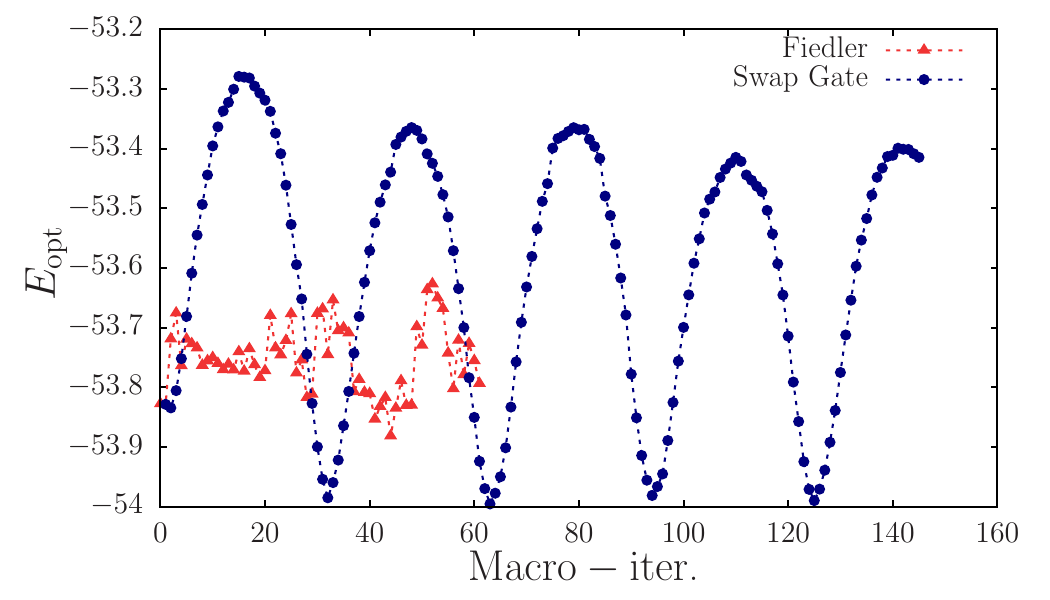}
    \caption{The ground state energy as a function of macro iteration number for the $8 \times 8$ Hubbard model \eqref{eq:ham-hubbard} at $U=+4t$ with $D_\mathrm{opt}=128$. The best energy of the Fiedler-method is $E^\mathrm{Fiedler}_{\mathrm{best}} = -53.882t$, while the swap-gate-based method reaches $E^\mathrm{swap}_{\mathrm{best}} = -53.995t$. 
    }  
    \label{suppfig:hubbard}
\end{figure*}

In the repulsive side, for $U=4$, because of the slower convergence of both methods, and also to demonstrate a hybrid solution, instead of initializing the Hamiltonian in the original, real-space modes, we start optimization from the best basis ($E_{\rm{opt,init}}^{\rm{Fiedler}}=-53.7623$) found by the heuristic protocol using Fiedler-vector-based reorderings~\cite{S.Menczer-2023d}. Here, the ground state shows antiferromagnetic correlations, i.e., the optimal modes are expected to be spin-dependent. Consequently, we allowed different rotation angles for the two spin components. As shown in Supp. Fig.~\ref{suppfig:hubbard}, the swap-gate based protocol can further optimize the basis and leads to significantly lower ground state energy
in the order of $10^{-1}$.
However, the effect of swap layers results in a strong undamped oscillation of period $N/2$, i.e., the local mode-optimization sweeps cannot fully recover the optimal basis once the optimum is approached, due to the simplified protocol used to calculate the global cost function without the back ordering of the orbitals (see main text). To prove the superiority of the new protocol, we have also repeated the analysis using the same parameters ($N_{\mathrm{sweep}}=25$ and $D_{\mathrm{opt}}=128$), but the Fiedler-vector-based reorderings. We observed spurious fluctuations caused by the Fiedler-vector-based reorderings, and the best energy ($E^{\mathrm{Fiedler}}_\mathrm{best} =-53.882t$) is also higher by 
$0.113t$ than $E^{\mathrm{swap}}_{\mathrm{best}} = -53.995t$.
The corresponding energy densities are $\epsilon^{\mathrm{Fiedler}}_{\mathrm{best}} = -0.8419$  and $\epsilon^{\mathrm{swap}}_{\mathrm{best}} = -0.8437$. These energy densities are measured at low bond dimension $M_\mathrm{opt}=128$. Lower results can be obtained by using an optimized basis and 
then increasing the bond dimension up to $ M\sim8000-10000$ (without further basis optimization). However, we note that the initial low-$M$ energy difference remains partly present in the large-$M$ calculations too~\cite{S.Menczer-2023d}; consequently, it is mandatory to optimize the basis as much as possible to gain significantly more accurate results for the large-scale simulations as well.  

\subsection{Strongly correlated molecular systems.}

\begin{figure*}
    \centering
    \includegraphics[width=0.78\textwidth]{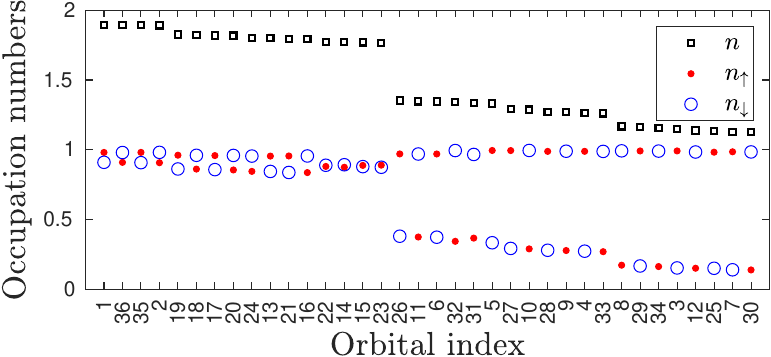}
    \caption{
    Occupation number profile obtained for the Fe$_4$S$_4$ in a CAS(54,36) model space using DMRG with $D=2048$ and the basis given in Ref.~\cite{S.Sharma-2014}. Labels in the x-axis indicate the corresponding orbital index.
    }  
    \label{suppfig:occup_fe4s4}
\end{figure*}

We close the numerical analysis of our new basis optimization protocol by providing benchmark calculations on the strongly correlated Fe$_4$S$_4$ molecular cluster~\cite{S.Sharma-2014}. In fact, this open shell system is identified and collected in a database maintained by IBM ~\cite{S.ibm-web} as a difficult problem for classical computers, where quantum advantage is expected to be gained via simulations on quantum computers. As further insights, we show in Fig.~\ref{suppfig:occup_fe4s4} the occupation number profile obtained by large-scale DMRG using $D=2048$ for the basis given in Ref.~\cite{S.Sharma-2014}. The twenty unpaired electrons of the four Fe atoms are clearly visible. The reference energy obtained with 4000 $\mathrm{SU}(2)$ multiplets, $D_{\mathrm{SU}(2)}=4000$, is -327.2396369~\cite{S.Sharma-2014}.

\begin{figure*}
    \centering
    \includegraphics[width=0.98\textwidth]{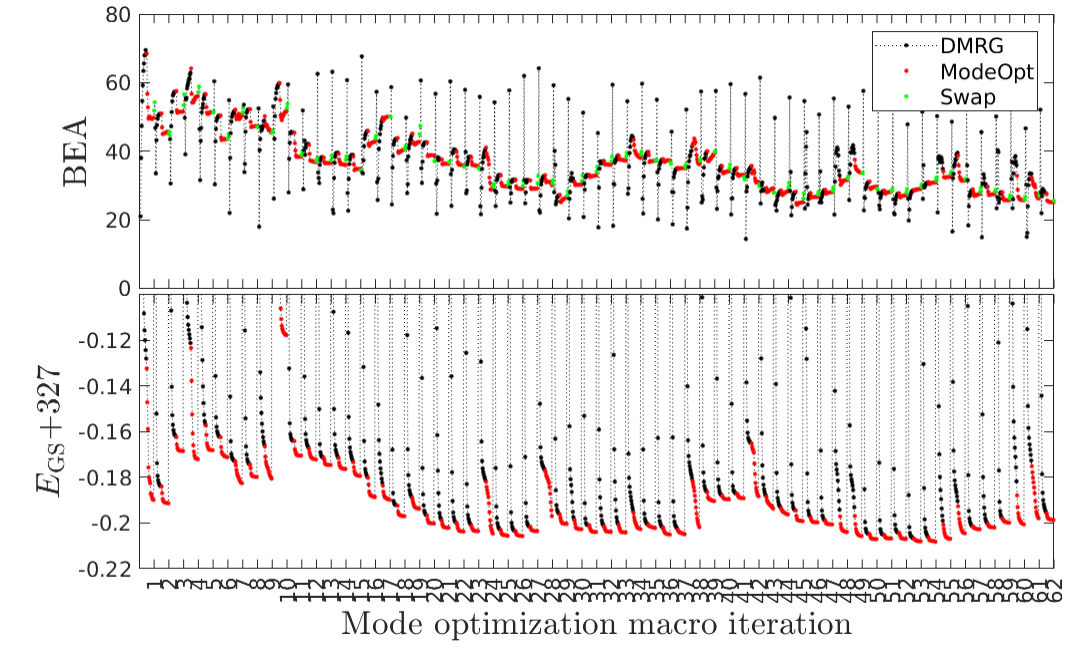}
    \caption{(a) Block entropy area $\calB_1(C)$ and (b) shifted ground state energy for the first 36 mode optimization macro iterations with $D_{\rm opt}=128$, $N_{\rm SW}^{\rm dmrg}=13$, $N_{\rm SW}^{\rm opt}=6$, and $N_{\rm SW}^{\rm opt,fin}=16$ 
    for the Fe$_4$S$_4$ in a CAS(54,36) model space~\cite{S.Sharma-2014}.
    Black dots and dashed lines indicate data points obtained via DMRG, red dots refer to local mode optimization and green dots to the application of swap gates.
    The end of each mode optimization macro cycle corresponds to every subsequent green symbol.
    }    
    \label{fig:bea_fe4s4}
\end{figure*}

In Fig.~\ref{fig:bea_fe4s4}(a) we show the Block entropy area $\calB_1(C)$ and (b) shifted ground state energy for the first 36 mode optimization macro iterations with $D_{\rm opt}=128$, $N_{\rm SW}^{\rm dmrg}=13$, $N_{\rm SW}^{\rm opt}=6$, and $N_{\rm SW}^{\rm opt,fin}=16$ as discussed in the main text.
Here, we allowed different rotation angles again for the two spin components.
Similarly to Fig.~4 in the main text
$\calB_1(C)$ decreases together with the ground state energy leading to $E_{\rm opt}=-327.208$. Without mode optimization using the energetic ordering and $D=128$ we obtained $E=-327.128$, while employing an optimized ordering, we obtained $E=-327.182$. 
Therefore, our mode-optimized energy is significantly lower than that one using the initial one-particle basis.

\begin{figure*}
    \centering
    \includegraphics[width=0.88\textwidth]{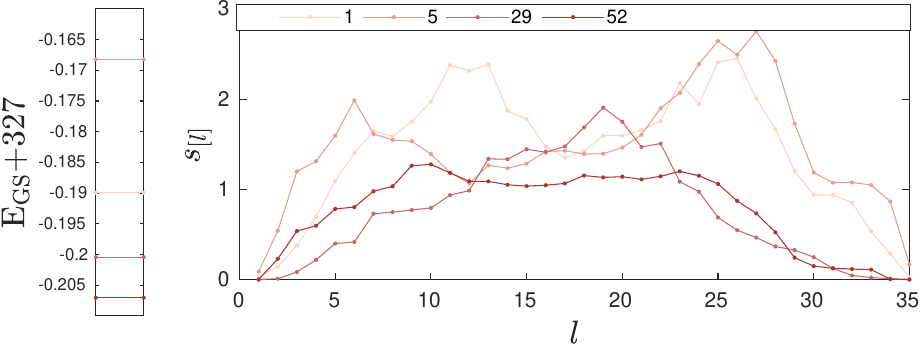}
    \caption{Ground state energy (left) and block entropy, $s_{[l]}$, as a function of the left block size, $\ell$ (right) for some selected mode transformation macro iteration cycles $1,5,29,52$, 
    with fixed bond dimension $D_{\rm mo}=128$ for the Fe$_4$S$_4$ cluster.
    The area under each curve corresponds to the block entropy area $\calB_1(C)$ as shown in Fig.~3 in the main text.
    }  
    \label{supfig:fe4s4_q1_S_swap}
\end{figure*}

In Fig.~\ref{supfig:fe4s4_q1_S_swap}, we show the block entropy profile for selected macro iteration steps. The tremendous reduction of BEA (from 49.8 to 25.1) is clearly visible. Moreover, via pure reordering the maximum of $s_{[l]}$ changes usually only slightly while while via mode optimization it drops from 2.8 to 1.2 in the current case. Note that the block entropy is a logarithmic function. Therefore, a large-scale DMRG calculation employing the optimized basis and fixing the truncation error or the entropy loss~\cite{S.Legeza-2003a, S.Legeza-2004} leads to a big reduction in the maximum bond dimension for a pre-defined accuracy threshold. These ultimately reduce wall time and speed up calculations, especially when GPU-accelerated supercomputers are employed ~\cite{S.Menczer-2024}. For further results and a more detailed analysis on quantum chemical systems, we guide interested readers to Ref.~\cite{S.Werner-2025}.   

\begin{figure*}
    \centering
    \includegraphics[width=0.76\textwidth]{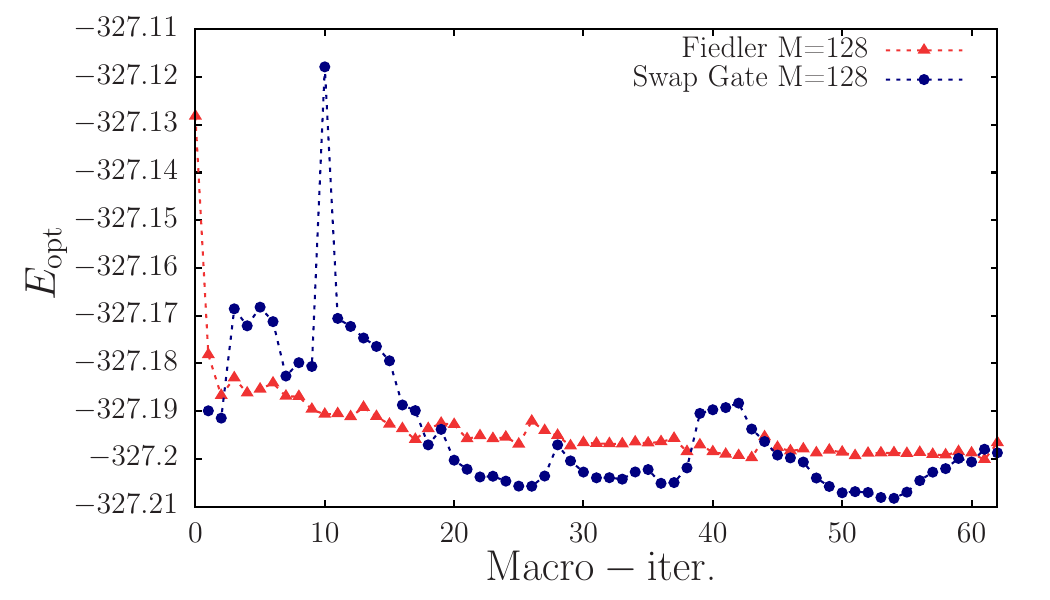}
    \caption{Similar to Fig.~\ref{suppfig:hubbard} but for the Fe$_4$S$_4$ molecular cluster.   
    Blue circles indicate the swap-gate-based, while red triangles are for the heuristic method. 
    }    
    \label{suppfig:fe4s4_fiedler_swap}
\end{figure*}

Finally, in Fig.~\ref{suppfig:fe4s4_fiedler_swap}, we show our results for the new swap-gate controlled mode optimization and the heuristic Fiedler-vector based protocol. Similar to Fig.~\ref{suppfig:hubbard}, both methods lower the energy significantly for a fixed bond dimension, but the swap-gate controlled protocol results in an optimized basis with significantly lower energy. The simplified calculation of the cost function results again in oscillation and a few scattered peaks, similarly to the results shown for the Hubbard model (Fig.~\ref{suppfig:hubbard}).   

%






\end{document}